\documentclass[12pt]{iopart}
 
\usepackage[german,english]{babel}
\usepackage{amssymb}
\usepackage{amsfonts}
\usepackage{graphics}

\begin{document}

\letter{Bubble dynamics in DNA}
 
\author{Andreas Hanke$^\dag$ and Ralf Metzler$^\sharp$}
\address{\dag Institut f{\"u}r Theoretische Physik, Universit{\"a}t
Stuttgart, Pfaffenwaldring 57, D-70550 Stuttgart, Germany}
\address{$\sharp$ NORDITA --- Nordic Institute for Theoretical Physics,
Blegdamsvej 17, DK-2100 Copenhagen {\O}, Denmark}
\ead{\mailto{hanke@theo2.physik.uni-stuttgart.de},
\mailto{metz@nordita.dk}}

\begin{abstract}
The formation of local denaturation zones (bubbles) in double-stranded
DNA is an important example for conformational changes of biological 
macromolecules. We study the dynamics of bubble formation in terms of 
a Fokker-Planck equation for the probability density to find a bubble 
of size $n$ base pairs at time $t$, on the basis of the free energy 
in the Poland-Scheraga model. Characteristic bubble closing and 
opening times can be determined from the corresponding first passage 
time problem, and are sensitive to the specific parameters entering 
the model. A multistate unzipping model with constant rates recently 
applied to DNA breathing dynamics 
[G. Altan-Bonnet \etal, Phys. Rev. Lett. {\bf 90}, 138101 (2003)] 
emerges as a limiting case.
\end{abstract}

\date{\today}
\submitto{\JPA}


\vspace*{0.8cm}
 
{\em Introduction.}
Under physiological conditions, the equilibrium structure of a DNA
molecule is the double-stranded Watson-Crick helix. At the
same time, in essentially all physiological processes involving DNA, 
for docking to the DNA, DNA binding proteins require access 
to the ``inside'' of the double-helix, and therefore the unzipping 
(denaturation) of a specific region of base pairs \cite{MBC,Revzin}. 
Examples
include the replication of DNA via DNA helicase and polymerase, and
transcription to single-stranded DNA via RNA polymerase. Thus,
double-stranded DNA has to open up locally to expose the otherwise 
satisfied bonds between complementary bases.

There are several mechanisms how such unzipping of double-stranded
DNA (dsDNA) can be accomplished. Under physiological conditions,
local unzipping occurs spontaneously
due to fluctuations, the {\it breathing} of dsDNA, which opens up
bubbles of a few tens of base pairs \cite{GKL87}. These 
breathing fluctuations may be supported by accessory 
proteins which bind to transient single-stranded regions, thereby 
lowering the DNA base pair stability \cite{Revzin}. 
Single molecule force spectroscopy opens the 
possibility to induce denaturation regions of controllable size, by 
{\it pulling} the DNA with optical tweezers \cite{WRB02}. In this way, 
the destabilising activity of the ssDNA-binding T4 gene 32 protein 
has been probed, and a kinetic barrier
for the single-strand binders identified \cite{PKW03}.
Denaturation bubbles can also be induced by {\it under-winding} 
the DNA double helix \cite{SABBC96}. A recent study of the dynamics 
of these twist-induced bubbles in a random DNA sequence shows that 
small bubbles (less than several tens of base pairs) are delocalised 
along the DNA, whereas larger bubbles become localised in AT-rich 
regions \cite{HMST03}. 
Finally, upon {\it heating\/}, dsDNA exhibits denaturation bubbles of 
increasing size and number, and eventually the two strands separate
altogether in a process called denaturation transition or melting 
\cite{PS,WB}. Depending on the specific sequence of the DNA molecule 
and the solvent conditions, the temperature $T_m$ at which one-half 
of the DNA has denatured typically ranges between $50^{\circ}$C and
$100^{\circ}$C. Due to the thermal lability of typical natural proteins,
thermal melting of DNA is less suited for the study of protein-DNA
interactions than force-induced melting  \cite{WRB02}. 
On the other hand, the controlled melting of DNA by heating is 
an important step of the PCR method for amplifying DNA samples 
\cite{Mullis}, with numerous applications in biotechnology 
\cite{SS03}. 

The study of the bubble dynamics in the above processes 
is of interest in view of a better understanding of the interaction 
with single-stranded DNA binding proteins. This interaction involves
an interplay between different time scales, e.g., the relaxation time
of the bubbles and the time needed for the proteins to rearrange 
sterically in order to bind \cite{karpel}. Dynamic probes such as
single-molecule force spectroscopy \cite{WRB02} and molecular 
beacon assays \cite{KB02} may therefore shed light on the 
underlying biochemistry of such processes.

In a recent experiment by Altan-Bonnet \etal \cite{BLK03}, 
the dynamics of a single bubble in dsDNA was measured by 
fluorescence correlation spectroscopy. 
It was found that in the breathing domain of the DNA construct
(a row of $18$ AT base pairs sealed by more stable GC base pairs)
fluctuation bubbles of size $2$ to $10$ base pairs are formed 
below the melting temperature $T_m$ of the AT breathing domain. 
The relaxation 
dynamics follows a multistate relaxation kinetics involving a 
distribution of bubble sizes and successive opening and closing of
base pairs.
The characteristic relaxation time scales were
estimated from the experiment to within the range of
$20$ to $100\,\mu$s. Also in reference \cite{BLK03}, a simple
master equation of stepwise zipping-unzipping with constant
rate coefficients was proposed to successfully describe the data
for the autocorrelation function, showing that indeed the bubble
dynamics is a multistate process. The latter was confirmed in a 
recent UV light absorption study of the denaturation of DNA 
oligomers \cite{zocchi}.

In this work, we establish a general framework to study the 
bubble dynamics of dsDNA by means of a Fokker-Planck equation,
based on the bubble free energy function. The latter allows one
to include microscopic interactions in a straightforward
fashion, such that our approach may serve as a testing ground 
for different models, as we show below.
In particular, it turns out that the phenomenological rate
equation approach, corresponding to a diffusion with constant drift
in the space of bubble size $n$ used by Altan-Bonnet
\etal \cite{BLK03} to fit their experimental data corresponds 
to a limiting case of our Fokker-Planck equation.
However, the inclusion of additional microscopic interactions 
in such a rate equation approach is not straightforward \cite{risken}.
In what follows, we first 
establish the bubble free energy within the Poland-Scheraga (PS) 
model of DNA denaturation \cite{PS,WB}, and then derive the 
Fokker-Planck equation to describe the bubble dynamics both 
below and at the melting temperature of dsDNA.

{\em Bubble free energy.}
In the PS model, energetic bonds in the double-stranded, helical 
regions of the DNA compete with the entropy gain from the far more 
flexible, single-stranded loops \cite{PS,WB}. 
The stability of the double helix originates mainly from stacking 
interactions between adjacent base pairs, aside from the Watson-Crick 
hydrogen bonds between bases. In addition, the positioning of bases 
for pairing out of a loop state gives rise to an entropic contribution.
The Gibbs free energy 
$G_{ij} = H_{ij} - T S_{ij}$ for the dissociation of two paired 
and stacked base pairs $i$ and $j$ has been measured, and is 
available in terms of the enthalpic and entropic contributions 
$H_{ij}$ and $S_{ij}$ \cite{Blake}.
In the following we consider a homopolymer for simplicity, as 
suitable for the AT breathing domain in reference \cite{BLK03}.
For an AT-homopolymer ($i = j = {\rm AT}$), the Gibbs 
free energy per base pair in units of $k_B T$ yields  as
$\gamma \equiv \beta G_{ii} / 2 = 0.6$ at $37^{\circ}$C
for standard salt conditions 
$(0.0745\,\mbox{M-Na}^+)$.
Similarly, for a GC-homopolymer one finds the higher value 
of $\gamma = 1.46$ at $37^{\circ}$C.
The condition $\gamma = 0$ defines the melting temperature 
$T_m$ \cite{Blake,GO03}, thus 
$T_m({\rm AT}) = 66.8^{\circ}$C and $T_m({\rm GC}) = 102.5^{\circ}$C
(we assume that $G_{ij}$ is linear in $T$,
cf.~reference \cite{RB99}).
Above $T_m$, $\gamma$ becomes negative. For given 
$\gamma = \gamma(T)$, the statistical weight for the 
dissociation of $n$ base pairs obtains as
\begin{equation}
\label{bond}
W(n)=\exp(- n \gamma).
\end{equation}
Additional contributions arise upon formation of a loop 
within dsDNA. Firstly, an initial energy barrier has to 
be overcome, which we denote as $\gamma_1$ in units of 
$k_B T$. From fitting melting curves to long DNA, 
$\gamma_1 \approx 10$ was obtained,
so that the statistical weight for the initiation of a loop 
(cooperativity parameter), $\sigma_0 = \exp(- \gamma_1)$, 
is of order $10^{-5}$ \cite{WB,Blake}.
As the energy to extend an existing loop by one base pair is
smaller than $k_B T$, DNA melts as large cooperative domains.
Below the melting temperature $T_m$, the bubbles become smaller, 
and long range interactions beyond nearest neighbours become 
more important. In this case, the probability of bubble formation 
is larger, and $\gamma_1$ ranges between 3 and 5,
thus $\sigma_0 \lesssim 0.05$ \cite{HMST03} 
(cf.~section 5 in \cite{WB}). 
According to reference \cite{GO03}, the smallness of 
$\sigma_0$ inhibits the recombination of complementary DNA 
strands with mutations, making recognition more selective.
Secondly, once a loop of $n$ base pairs has formed, there 
is a weight $f(n)$ of mainly entropic origin, to be
detailed below. The additional weight of a loop of $n$ open 
base pairs is thus
\begin{equation}
\label{loopgen}
\Omega(n)=\sigma_0 \, f(n) \, .
\end{equation}
For large bubbles one usually assumes the form
$f(n) = (n+1)^{-c}$ \cite{WB,Blake}. Here, the value of the exponent 
$c=1.76$ corresponds to a self-avoiding, flexible ring
\cite{PS,WB,fisher}, which is classically used in denaturation 
modelling within the PS approach. Recently, the PS model has been 
considered in view of the order of the denaturation phase transition 
\cite{KMP00,GMO01,COS02,HM03,RG03}. 
Reference \cite{zocchi} finds by finite size scaling analysis 
of measured melting curves of DNA oligomers that the transition
is of second order. In reference 
\cite{KMP00}, the value $c=2.12$ was suggested, compare the 
discussion in references \cite{HMST03,HM03,GO03}.
For smaller bubbles (in the range of 1 to a few tens of base pairs), 
the appropriate form of $f(n)$ is more involved. 
In particular, $f(n)$ will depend on the finite persistence 
length of ssDNA (about eight bases), on the specific base 
sequence, and possibly on interactions between dissolved but 
close-by base pairs (cf.~section 2.1.3 in \cite{WB}). 
Therefore, the knowledge of $f(n)$ provides information 
on these microscopic interactions. 

For simplicity, we here adopt the simple form $f(n) = (n+1)^{-c}$ 
for all $n > 0$, and consider the loop weight
\begin{equation}
\label{loop}
\Omega(n)=\sigma_0 (n+1)^{-c} \, .
\end{equation}
We show that at the melting temperature the results for 
the relaxation times for the bubbles are different for the 
available values $c=1.76$ and $c=2.12$ quoted above.
This shows that the specific form of $f(n)$ indeed 
modifies experimentally accessible features of the 
bubble dynamics \cite{BLK03}.

\begin{figure}[h]
\begin{center}
\includegraphics{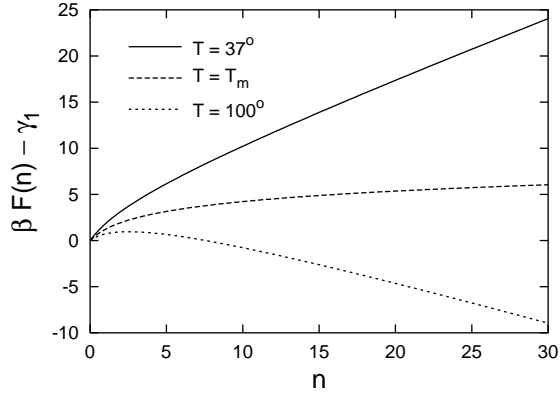}
\end{center}
\caption{Variable part of the bubble free energy (\protect\ref{free}), 
$\beta {\cal F}(n) - \gamma_1 = n \gamma(T) + c \ln(n+1)$, as a function 
of the bubble size $n$ for
$T=37^{\circ}$C ($\gamma\approx 0.6$), $T_m=66^{\circ}$C ($\gamma=0$), 
and $T=100^{\circ}$C ($\gamma\approx -0.5$). In the
latter case, a small barrier precedes the negative drift towards bubble
opening as dominated by the $\gamma<0$ contribution.
\label{fig1}}
\end{figure}

In what follows, we focus on a single bubble in dsDNA,
neglecting its interaction with other bubbles. Since due
to $\sigma_0 \ll 1$ the mean distance between bubbles 
($\sim 1 / \sigma_0$ \cite{PS}) is large,
this approximation is justified as long as the bubbles 
are not too large \cite{HMST03}. It also corresponds to the 
situation studied in the recent experiment by Altan-Bonnet 
\etal \cite{BLK03}. According to the above, the total free 
energy ${\cal F}(n)$ of a single bubble with $n>0$ open base 
pairs follows in the form
\begin{equation}
\label{free}
\beta {\cal F}(n) = - \ln \left[ W(n) \Omega(n) \right]
= n \gamma(T) + \gamma_1 + c \ln(n+1) \, ,
\end{equation}
where the dependence on the temperature $T$ enters only via 
$\gamma = \gamma(T)$. We show the free energy (\ref{free}) 
in figure \ref{fig1} for $c=1.76$ and for the parameters of an
AT-homopolymer, for physiological temperature $T=37^{\circ}$C,
at the melting temperature $T_m=66^{\circ}$C, and at
$T=100^{\circ}$, compare the discussion below.

{\em Bubble dynamics.}
In the generally accepted multistate unzipping model, the double strand
opens by successively breaking Watson-Crick bonds, like opening a zipper
\cite{nelson,kittel}. As $\gamma$ becomes small on increasing the temperature,
thermal fluctuations become relevant and cause a random walk-like propagation
of the zipper locations at both ends of the bubble-helix joints.
The fluctuations of the bubble size can therefore be described in the
continuum limit through a
Fokker-Planck equation for the probability density function (PDF)
$P(n,t)$ to find at time $t$ a bubble consisting of $n$ denatured base
pairs, following a similar reasoning as pursued in the modelling of the dynamics
of biopolymer translocation
through a narrow membrane pore \cite{translocation}. To establish this
Fokker-Planck equation, we combine the continuity equation (compare reference
\cite{translocation})
\begin{equation}
\label{conti}
\frac{\partial P(n,t)}{\partial t}+\frac{\partial j(n,t)}{\partial n}=0
\end{equation}
with the expression for the corresponding flux,
\begin{equation}
\label{flux}
j(n,t)=-D\left(\frac{\partial P(n,t)}{\partial n}+\frac{P(n,t)}{k_BT}
\frac{\partial {\cal F}}{\partial n}\right)
\end{equation}
where it is assumed that the potential exerting the drift is given by the
bubble free energy (\ref{free}). In equation (\ref{flux}), we incorporated
an Einstein relation of the form $D=k_BT\mu$, where 

\noindent
the mobility $\mu$ has
dimensions $[\mu]={\rm sec}/({\rm g}\cdot{\rm cm}^2)$, and therefore $[D]={\rm
sec}^{-1}$ represents an inverse time scale. By combination of equations
(\ref{free}), (\ref{conti}) and (\ref{flux}), we retrieve the Fokker-Planck
equation for $P(n,t)$ 
\footnote{Note that the operator $\frac{\partial}{\partial n}$
acts also on $P(n,t)$.}
\begin{equation}
\label{fpe}
\frac{\partial P(n,t)}{\partial t}=D\left(\frac{\partial}{\partial n}\left\{
\gamma+\frac{c}{n+1}\right\}+\frac{\partial^2}{\partial n^2}\right)P(n,t).
\end{equation}
Thus, we arrived at a reduced 1D description of the bubble dynamics
in a homopolymer, with the bubble size $n$ as the effective `reaction' 
coordinate. For a heteropolymer, also the position of the 
bubble within the double helix, i.e., the index $m$ of the first open 
base pair, becomes relevant. In this case, the bubble free energy 
${\cal F}$ and thus the PDF depend on both $m$ and $n$, and on the 
specific base sequence; the corresponding generalisation of equation 
(\ref{fpe}) is straightforward. 
In a random sequence, additional phenomena may occur, 
such as localisation of larger bubbles \cite{HMST03}. 
Finally, to establish the Fokker-Planck equation (\ref{fpe}), we assume 
that changes of the bubble size $n$ occur slower than other degrees of 
freedom of the PS free energy within the bubble region (e.g., Rouse-Zimm
modes). This assumption seems reasonable due to the long bubble dynamics' 
relaxation time scales of $20$ to $100\,\mu$s \cite{BLK03}, and the 
good approximation of bubble independence \cite{HMST03}. 

By rescaling time according to $t\to Dt$, 
the Fokker-Planck equation (\ref{fpe})
can be made dimensionless, a representation we are going to use in the
numerical evaluation below. The formulation in terms of a Fokker-Planck
equation makes it possible to derive the characteristic times for bubble
closing and opening in terms of a first passage time problem. That is,
for bubble
closing, the associated mean closing time follows as the mean first passage
time to reach bubble size $n=0$ after starting from the initial bubble size
$n_0$. We now determine these characteristic times for the three regimes
defined by $\gamma$ with respect to temperature $T$.

\begin{figure}[h]
\begin{center}
\includegraphics{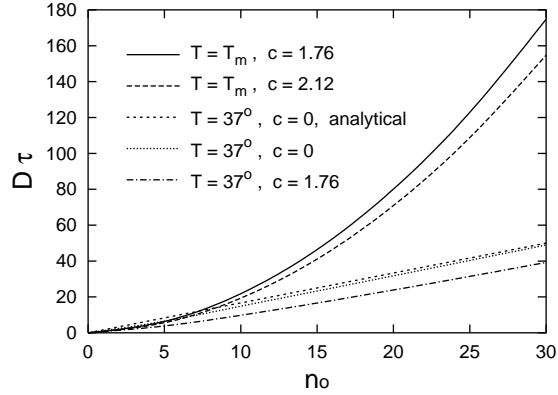}
\end{center}
\caption{Characteristic bubble closing times $\tau$ 
as a function of initial bubble size $n_0$ for an AT-homopolymer,
obtained from the Fokker-Planck equation (\protect\ref{fpe})
by numerical integration.
At $T=37^{\circ}$C, the result for $c = 1.76$ is compared to 
the approximation $c=0$ which leads to somewhat larger closing times. 
The analytical solution for $c=0$ compares well with the numerical 
result, the slight discrepancy being due to the reflecting boundary 
condition applied in the numerics, in comparison to the natural 
boundary condition at $n\to\infty$ used to derive equation 
(\protect\ref{mfpt}). At the melting temperature $T_m=66^{\circ}$C, 
the closing times for the values $c=1.76$ and $c=2.12$ can be 
distinguished.
\label{fig2}}
\end{figure}
(i) $T<T_m$. In this regime, the drift consists of two contributions, the
constant drift $D\gamma$ and the loop closure component $Dc/(n+1)$, which
decreases with $n$. For large $n$, we can therefore approximate the drift
by the constant term $D\gamma$, and in this limit the Fokker-Planck
equation (\ref{fpe}) is equivalent to the continuum limit of the master
equation used in reference \cite{BLK03} to describe the experimental
bubble data. In particular, we can identify our two independent parameters
$D$ and $\gamma$ with the rate constants $k_+$ and $k_-$ to open and close
a base pair introduced in \cite{BLK03}, respectively: $D\equiv (k_++k_-)/2$ and
$\gamma\equiv 2(k_--k_+)/(k_++k_-)$. In this approximation, the correlation
functions used successfully to fit the experimental results in \cite{BLK03}
can be derived from the Fokker-Planck equation (\ref{fpe}). Moreover, we
can deduce the mean first passage time PDF for
a bubble of initial size $n_0$ to close in the exact analytical form
(compare \cite{redner})
\begin{equation}
\label{mfpt}
f(0,t)=\frac{n_0}{\sqrt{4\pi Dt^3}}\exp\left\{-\frac{(n_0-D\gamma t)^2}{
4D t}
\right\},
\end{equation}
which decays exponentially for large $n$. In particular, from (\ref{mfpt})
the characteristic (mean) first passage time for bubble closing,
\begin{equation}
\label{close}
\tau=n_0/(D\gamma)
\end{equation}
follows, which is linear in the initial bubble size $n_0$. In figure
\ref{fig2}, we compare this analytical result for the value $\gamma(37
^{\circ}\mbox{C})$ with the characteristic closing times using the full
drift term from equation (\ref{fpe}), obtained from numerical integration.
It can be seen that the qualitative behaviour for both cases with and
without the $Dc/(n+1)$ term is very similar, but that in the presence of
the loop closure component the characteristic times are reduced.

(ii.) $T=T_m$. At the melting temperature, the drift in equation
(\ref{fpe}) is solely given by the loop closure term $Dc/(n+1)$. In
figure \ref{fig2}, we plot the characteristic bubble closing time
obtained numerically. In comparison to the above case $T<T_m$, the
faster than linear increase as function of initial bubble size $n_0$
is distinct. Keeping in mind that $Dc/(n+1)$ becomes very small for increasing
$n$, this behaviour can be qualitatively understood from the approximation
in terms of a drift-free diffusion in a box of size $n_0$, in which the
initial condition $P(n,0)=\delta(n-n_0)$ is located at a reflecting
boundary, and at $n=0$ an absorbing boundary is placed. This problem 
can be solved analytically, with the result $\tau=n_0^2/(2D)$ for the 
characteristic escape time 
\footnote{This result can, for instance, be obtained through
the method of images, compare reference \protect\cite{bvp}.}
in which the quadratic dependence on $n_0$ contrasts the linear 
behaviour in the result (\ref{close}). 
Thus, at the melting temperature $T_m$, the tendency for a
bubble to close becomes increasingly weaker for larger bubble sizes, 
and therefore much larger bubbles can be formed, in contrast to the 
case $T<T_m$. 
A further comparison to the value $c=2.12$ mentioned above demonstrates that
a clear quantitative difference in the associated closing times exists, see
figure \ref{fig2}. However, the qualitative behaviour remains unchanged. In
principle, the study of the bubble dynamics can therefore be used to discern
different models for the loop closure factor.

(iii.) $T>T_m$. Above the melting temperature, the drift is governed by
the interplay between the loop closure component $Dc/(n+1)$ tending to close
the bubble, and the bubble free energy $D\gamma(T>T_m)<0$, which causes
a bias towards bubble opening. In figure \ref{fig1}, we show for the
AT-homopolymer case how the overall drift potential after a
small initial activation barrier becomes negative, and the dynamics is
essentially governed by the $\gamma$-contribution. As a consequence, from
the result (\ref{mfpt}), we find that the associated mean first passage
time diverges, i.e., the bubble on average increases in size until the
entire DNA is denatured, as expected from the PS model.

By symmetry, similar results hold for the bubble opening process.
However, the existence of the bubble initiation energy $\gamma_1$
involves an additional Arrhenius factor, which is not included in
the Fokker-Planck equation (\ref{fpe}), and which reduces the opening
probability, causing an increase of the associated opening time,
cf.~reference \cite{BLK03}.

{\em Conclusions.}
From the bubble free energy of a single, independent bubble in the
Poland-Scheraga theory of DNA melting, we derived a Fokker-Planck equation
for the PDF to find a bubble created by denaturation of $n$
base pairs at time $t$. This formulation allows for the calculation of the
characteristic times scales for bubble closing and opening in terms of first
passage time problems. Three different regimes were distinguished: (i) below
the melting temperature, the characteristic bubble closing time increases
linearly in bubble size, and the drift can be approximated by the constant
value $D\gamma$. In this approximation, the Fokker-Planck equation matches
the continuum version of the master equation employed previously in an
experimental study of DNA breathing \cite{BLK03}. (ii) At the melting 
temperature, an approximately quadratic growth of the bubble closing time
as a function of bubble size is observed, which can be explained by noting
that the $1/(n+1)$-dependence of the loop closure drift component can be
neglected for larger $n$, leading to pure diffusion.
In this approximation, the exact analytical results
indeed lead to the quadratic dependence observed in the numerical results.
(iii) Above the melting temperature, the characteristic closing time in our
model diverges, consistent with the fact that on average the DNA follows the
trend towards the thermodynamically favourable state of complete denaturation.

The expression for the Gibbs free energy used in our approach involves a purely
entropic contribution for a single-stranded bubble. It was suggested in
reference \cite{BLK03} that also in denaturation bubbles a residual stacking
energy $\varepsilon_s$ is present. In this case, the value of $\gamma$ would
have to be corrected by this $\varepsilon_s$.

From the Fokker-Planck equation (\ref{fpe}), in which the microscopic
interactions enter via the free energy (\ref{free}), one can 
derive measurable quantities such as moments and dynamic correlation 
functions \cite{risken}.
In principle, the Fokker-Planck equation involves one free
parameter, the time scale $1/D$, while the values for the other parameters
are known. However, by fitting sufficiently accurate experimental data for
DNA bubble dynamics at different temperatures, the values for additional
parameters may be extracted by help of our general Fokker-Planck framework 
probing suitable free energy functions. 

\ack

We are happy to acknowledge helpful discussions with Terence Hwa,
Yacov Kantor, Mehran Kardar, Richard Karpel, Udo Seifert, and Mark 
Williams. We also thank Oleg Krichevsky for sending us a preprint 
of reference \cite{BLK03} prior to publication.

\end{document}